\documentclass[aps,prl,twocolumn,superscriptaddress,nofootinbib,showpacs,letterpaper]{revtex4}

\usepackage{showlabels}
\usepackage{verbatim}
\usepackage[usenames,dvipsnames]{color}
\usepackage[pdftex]{graphicx}
\usepackage[english]{babel}
\usepackage{amsmath,amssymb}
\usepackage[colorlinks=true,citecolor=blue]{hyperref}
\usepackage{bm}

\begin{document}

\newcommand{\be}{\begin{equation}}
\newcommand{\ee}{\end{equation}}
\newcommand{\R}[1]{\textcolor{red}{#1}}
\newcommand{\mycomment}[1]{}
\newcommand{\here}{\textcolor{red}{$\square$}}

\title{Macroscopic Quantum Mechanics in a Classical Spacetime}
\author{Huan Yang}

\author{Haixing Miao}
\affiliation{Theoretical Astrophysics 350-17, California Institute of Technology, Pasadena, California 91125, USA}

\author{Da-Shin Lee}
\affiliation{Department of Physics, National Dong Hwa University, Hua-Lien, Taiwan 974, Republic of China}
\affiliation{Theoretical Astrophysics 350-17, California Institute of Technology, Pasadena, California 91125, USA}

\author{Bassam Helou}
\affiliation{Theoretical Astrophysics 350-17, California Institute of Technology, Pasadena, California 91125, USA}

\author{Yanbei Chen}
\affiliation{Theoretical Astrophysics 350-17, California Institute of Technology, Pasadena, California 91125, USA}

\begin{abstract}
We apply the many-particle Schr\"{o}dinger-Newton equation, which describes the coevolution
of a many-particle quantum wave function and a classical space-time geometry, to macroscopic mechanical objects.
By averaging over motions of the objects' internal degrees of freedom, we obtain an effective Schr\"{o}dinger-Newton equation for their centers of mass, which can be monitored and manipulated at quantum levels by state-of-the-art optomechanics experiments.
For a single macroscopic object moving quantum mechanically within a harmonic potential well, its quantum uncertainty is found to evolve at a  frequency different from its classical eigenfrequency --- with a difference that depends on the internal structure of the object --- and can be observable using current technology.
For several objects, the  Schr\"odinger-Newton equation predicts semiclassical motions just like Newtonian physics, yet quantum uncertainty cannot be transferred from one object to another. 
\end{abstract}
\maketitle

\begin{figure}[]
\includegraphics[width=0.35\textwidth]{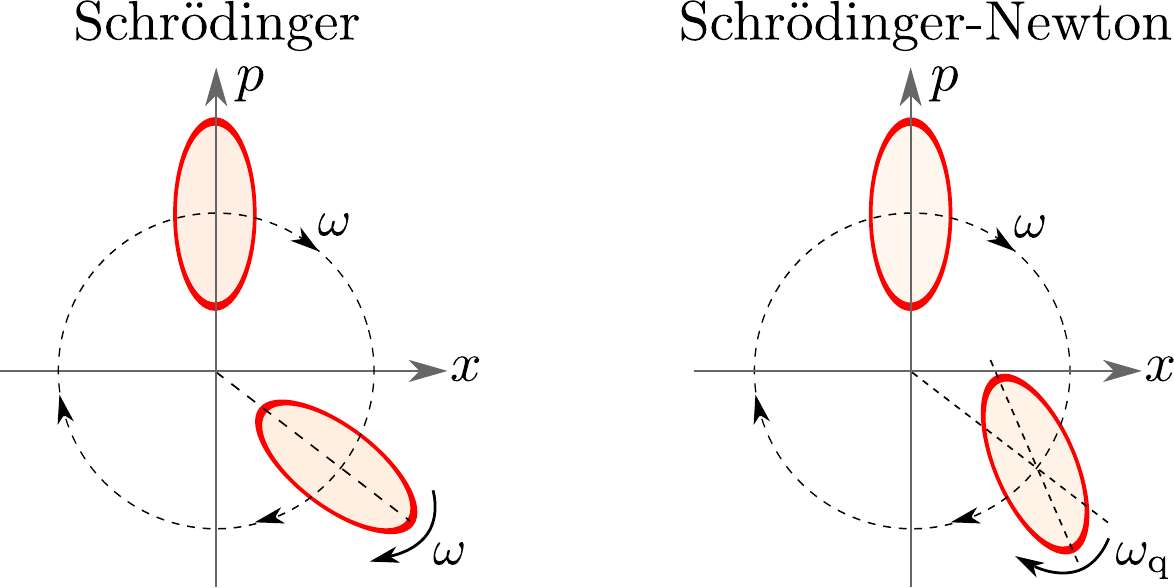}
\caption{(Color online). Left Panel: according to standard quantum mechanics, both the vector $(\langle x\rangle,\langle p\rangle)$ and the uncertainty ellipse of a Gaussian state for the CM of a macroscopic object rotate clockwise in phase space, at the same frequency $\omega =\omega_{\rm CM}$.
Right panel:  according to the CM Schr\"odinger-Newton Equation~\eqref{eqreal}, $(\langle x\rangle,\langle p\rangle)$ still rotates at $\omega_{\rm CM}$, but the uncertainty ellipse rotates at $\omega_{\rm q} \equiv ({\omega_{\rm CM}^2+\omega_{\rm SN}^2})^{1/2}>\omega_{\rm CM}$.
 \label{fig:2freq}}
\end{figure}

\noindent {\it Introduction and summary.---} Testing non-relativistic quantum mechanics using macroscopic objects has has been a minor approach towards the search for effects of quantum gravity. 
Apart from the standard formulation of linearized quantum gravity~\cite{Bonifacio}, which seems rather
implausible to test in the lab, several signatures have been conjectured: (i) {gravity decoherence}~\cite{diosi2, diosi3,diosi4,penrose1,penrose2,Marshall,Hong,jasper,kaltenbaek,romero1,romero2}, where gravity introduces decoherence to macroscopic quantum superpositions; (ii) modifications to canonical quantization motivated by the existence of a minimum length scale~\cite{maggiore,das,pikovski}; and (iii) { semiclassical gravity}~\cite{moller,rosenfeld,carlip}, which will be the subject of this paper. As originally suggested by
M{\o}ller \cite{moller} and Rosenfeld \cite{rosenfeld}, spacetime structure might {\it still remain classical} even if it
is sourced by matters of quantum nature, if we impose
 ($G=c=1$)
\begin{equation}\label{eqintroavg}
G_{\mu\nu}=8\pi  \langle \psi| \hat{T}_{\mu\nu} |\psi\rangle\,.
\end{equation}
Here $G_{\mu\nu}$ is the Einstein tensor of a (3+1)-dimensional classical spacetime,
$\hat T_{\mu\nu}$ is the operator for the energy-stress tensor, and $|\psi(t)\rangle$ is the wave
function of all matters that evolve within this classical spacetime. \here

Many arguments exist {\it against} semiclassical gravity. Some
rely on the conviction that a classical system cannot properly integrate with a
quantum system without creating contradictions. Others are based on ``intrinsic'' mathematical inconsistencies, the most famous one between Eq.~\eqref{eqintroavg}, state collapse, and $\nabla^\nu G_{\mu\nu}=0$~\cite{Wald}.  Towards the former argument, it is the aim of this paper to explicitly work out the effects of classical gravity on the quantum mechanics of macroscopic objects; although we will find them counter intuitive, they do not seem dismissible right away. In fact, we shall find these effects ``right on the horizon of testability'' by current experimental technology. Towards the latter argument, we shall remain open minded regarding the possibility of getting rid of quantum state reduction while at the same time avoiding the many-world interpretation of quantum mechanics~\cite{Everett,page} (also see Supplementary Material).

The non-relativistic version of  Eq.\,\eqref{eqintroavg}, the so-called Schr\"{o}dinger-Newton (SN)
equation, has been extensively studied for single particles~\cite{diosi1,Moroz,Harrison,salzman,adler,meter,Guzman}.  In this paper, we  consider instead a macroscopic object consisting many particles, and will show that within
certain parameter regimes, the center-of-mass (CM) wavefunction approximately satisfies
the following SN equation:
\begin{equation}\label{eqreal}
   i\hbar \frac{\partial \Psi}{\partial t}\!=\! \left [-\frac{\hbar^2\nabla^2}{2M}+\frac{1}{2}M \omega_{\rm CM}^2 x^2
   + \frac{1}{2} \mathcal{C}
   (x-\langle x\rangle)^2\right ]\Psi.
\end{equation}
Here $\langle x\rangle\equiv\langle \Psi|\hat x |\Psi\rangle$ is the expectation
value of CM position; $\omega_{\rm CM}$ is the eigenfrequency in absence of gravity,
determined by how the CM is confined; $\mathcal{C}$ is the SN coupling constant,
from which we introduce
$\omega_{\rm SN} \equiv \sqrt{\mathcal{C}/M}$.
%Far below the Debey tempreature, $\mathcal{C}$ for a piece of crystal is given by $\omega_{\rm SN}^2 =   Gm/(12\sqrt{\pi}\Delta x_{\rm zp}^3)$
%with $\Delta x_{\rm zp}$ the zero-point position uncertainty of atoms.
For Si crystal at 10\,K, we estimate $\omega_{\rm SN}\sim 0.036\,\rm s^{-1}$, much larger than the naively expected  $ \sqrt{G \rho_0}$ from the object's mean density $\rho_0$, due to the high concentration of mass near lattice points.

For a single macroscopic object prepared in a squeezed Gaussian state, Eq.~\eqref{eqreal} leads to different evolutions of expectation values and quantum uncertainties, as illustrated in Fig.\,\ref{fig:2freq}.
Such a  deviation can be  tested by optomechanical devices in the quantum
regime\,\cite{optomechanics_review,oconnel,qfproduct,teufel,Purdy}.   For two macroscopic objects interacting through gravity, we  show further,  using the two-body counterpart of Eq.~\eqref{eqreal}, that classical gravity cannot be used to transfer quantum uncertainties --- experimental demonstration of this effect will be much more difficult than demonstrating modifications in single-object dynamics.

We emphasize that it is {\it not} our aim to use the SN equation to explain the collapse of quantum states, or to provide a pointer basis for gravity decoherence, as has been attempted in the literature~\cite{diosi1,Moroz,Harrison,salzman,adler,meter,Guzman}.   We will take a conservative strategy, avoiding  experimental regimes with exotic wavefunctions~\cite{kaltenbaek,romero1,romero2}, and restraining ourselves to Gaussian states whose evolutions deviate little from predictions of standard quantum mechanics: just enough  to be picked up by precision measurements.   In this way, solutions to SN equation we consider are much less complex than those in previous literature~\cite{diosi1,Moroz,Harrison,salzman,adler,meter,Guzman}.

\noindent {\it Many-particle SN equation.---} For $n$ non-relativistic particles, if we denote their joint wave function as
$\varphi(t,\,\mathbf{X})$ with $3n$-D vector ${\mathbf{X}}\equiv ({\mathbf{x}}_1, \,\cdots,\, {\mathbf{x}}_n)$ and ${\mathbf{x}}_k$ the 3-D spatial coordinate of $k$-th particle, then the many-particle SN equation, obtained by Diosi and Penrose~\cite{diosi1,penrose1}, is
\be\label{eqSN}
i\hbar{\partial}_t \varphi=\sum_k \left[-\frac{\hbar^2\nabla^2_k}{2m_k}
+\frac{m_k\,U(t, \mathbf{x}_k)}{2}\right] \varphi+ V(\mathbf{X})\varphi\,,
\ee
where $V(\mathbf{X})$ is potential energy for non-gravitational interactions,
while the Newtonian potential $U$ is given by
\be\label{eqU}
\nabla^2 U(t,\,\mathbf{x}) =4\pi\sum_j \int {\rm d}^{3n}{\mathbf{X}} \, |\varphi(t, \mathbf{X})|^2 m_j\,\delta(\mathbf{x}-\mathbf{x}_j)\,.
\ee

\noindent{\it The Center of Mass and the Separation of Scales.---}
Equations~\eqref{eqSN} and \eqref{eqU} are still not suitable for experimental studies, because we cannot separately access each particle in a macroscopic object.
In optomechanical devices, a light beam often probes (hence acts back onto) the average displacements of atoms within the first few layers of the reflective coating of a mirror-endowed mass. Motion of this effective surface  can often be well-approximated by the CM motion of the entire object (see \cite{oconnel,teufel,corbitt});  the error of this approximation is referred to as the
``internal thermal noise'', and has been shown to be suppressible below the free-mass Standard Quantum Limit (SQL)~\cite{BK}, a quantum level of CM motion defined by the object's total mass and the measurement time scale~\cite{coating_thermal_noise}.  This suppression is possible because: (i) we tend to measure CM motion by averaging over a large number of atoms at the surface of the object, and (ii) we measure CM motion over a time scale much longer than ones at which atoms oscillate due to thermal or zero-point fluctuations.   Obtaining the SN equation for the CM is therefore central to the experimental test of this model.  Before doing so,  let us consider the separation of temporal and spatial scales in the motion of a macroscopic piece of crystal.
%
%\smallskip
%\noindent {\it Separation of scales.---}To be specific, let us consider a macroscopic object
%(e.g., a crystal) with equal-mass particles (atoms, mass $m$) positioned on a uniform 3-D lattice.
%Before actually deriving the SN equation for the CM, let us survey the time and length
%scales in our system, which are critical for separating the CM motion from the
%internal motions---the oscillations of the particles around their equilibrium positions on the lattice.
%

The scales of CM motion are determined externally by how we confine the object during measurement, and by how we measure it.
Here we consider motions with $\omega_{\rm CM}/(2\pi)$ from Hz to kHz scale. If thermal noise level is below the free-mass SQL~\cite{BK}, then one can either use optical or feedback trapping to create mechanical oscillators with coherence time $\tau_{\rm CM}$ longer than $1/\omega_{\rm CM}$~\cite{helge,haixing}.  Although not yet achieved, research towards sub-SQL devices in the Hz -- kHz regime is being actively pursued~\cite{thomas,LIGO3,corbitt}. In this regime, we have $\Delta x_{\rm CM} \sim \sqrt{\hbar /(M\omega_{\rm CM})}$; for 1\,g$<M<10\,$kg,  $\Delta x_{\rm CM} \sim 10^{-19}$--$10^{-17}\,$m.

By contrast, the internal motions of atoms are due to excitation of phonons~\cite{debye}, with a total variance of~\cite{BF}
\begin{equation}
\langle x^2\rangle \equiv \frac{B^2}{8\pi^2} = \frac{\hbar^2}{mk_B T}\int_0^{+\infty} \frac{g(\nu) }{\xi}\left(\frac{1}{2}+\frac{1}{e^\xi -1}\right) d\nu
\end{equation}
where $B$ is also known as the ``B-factor'' in X-ray diffraction, $\xi=h\nu/k_B T$, $g(\nu)$ is the phonon density of states; first term in the bracket gives rise to zero-point uncertainty $\Delta x_{\rm zp}^2$, while the second gives rise to thermal uncertainty $\Delta x_{\rm th}^2$. These have been studied experimentally by X-ray diffraction, through measurements of the Debye-Waller factor~\cite{DWF},  and modeled precisely  (for Si crystal, see Ref.~\cite{Flensburg}).
Much below the Debye temperature, one can reach: $\Delta x_{\rm th} \ll  \Delta x_{\rm zp}$, with most atomic motion due to zero-point fluctuations near the Debye frequency $\omega_{\rm D}$. For Si crystal, $\omega_{\rm D}\sim 10^{14}\,\mathrm{s}^{-1}$, $\Delta x_{\rm zp} = 4.86\times 10^{-12}\,$m, and  $\Delta x_{\rm th}(293\,\mathrm{K}) = 5.78 \times 10^{-12}\,$m~\cite{Flensburg}. At lower temperatures, $\Delta x_{\rm th}\propto T$, therefore on the scale of $\sim 10\,$K, at which our proposed experiment operates, we have $\Delta x_{\rm zp} \gg \Delta x_{\rm th} \gg \Delta x_{\rm CM}$.

\noindent {\it SN equation for the CM.---}  For a crystal with $n$ atoms,
the CM is at ${\mathbf{x}}_{\rm CM}=({1}/{n})\sum_k {\mathbf{x}}_k$, motion of the $k$-th atom in CM frame is
$\mathbf{y}_k\equiv \mathbf{x}_k -\mathbf{x}_{\rm CM}$. In standard quantum mechanics, for inter-atom interaction that only depends on the separation
of atoms, the CM and internal DOFs are separable:
$\varphi(t,\mathbf{X}) =\Psi_{\rm CM}(t,\mathbf{x}) \Psi_{\rm int}(t,\,\mathbf{Y})$,
with $3(n-1)$-D vector $\mathbf{Y}\equiv(\mathbf{y}_1, \,\cdots, \, \mathbf{y}_{n-1})$. The two wavefunctions evolve independently:
\begin{align}
i\hbar\partial_t \Psi_{\rm CM} (t,\mathbf{x}) &= H_{\rm CM}\Psi_{\rm CM} (t,\mathbf{x}) \,,\\
i\hbar\partial_t \Psi_{\rm int}(t,\mathbf{Y}) &=H_{\rm int} \Psi_{\rm int}(t,{\mathbf{Y}})\,.
\end{align}
For classical gravity, let us first still assume separability: $\varphi =\Psi_{\rm CM}\Psi_{\rm int}$, and we will
show this remains true (with negligible error) under evolution. Specifically, sum of SN terms in Eq.~\eqref{eqSN} becomes
%
%
%In the case of the SN equation, the Newtonian potential $U$ in Eq.\,\eqref{eqSN}
%does not allow a strictly separable evolution of $\Psi_{\rm CM}$ and $\Psi_{\rm int}$,
\begin{align}
V_{\rm SN}(\mathbf{x},\mathbf{Y})= &\sum_{k}{m_kU(\mathbf{x}_k)}/{2} \nonumber\\
%=& -\frac{Gm^2}{2}\sum_{j,k} \int \frac{|\Psi_{\rm CM}^2(\mathbf{z})| |\Psi^2_{\rm int} (\mathbf{Y'})| }{|\mathbf{x}+ \mathbf{y}_j -\mathbf{y}_k' -\mathbf{z}|}d^{3n-3}\mathbf{Y'} d^3\mathbf{z} \nonumber\\
=&\sum_k\int \varepsilon\left[\mathbf{x} - \mathbf{z} +\mathbf{y}_k\right]\Psi_{\rm CM}^2(\mathbf{z})  d^3\mathbf{z}\,.
\end{align}
Here we have suppressed dependence on time and defined
\begin{align}
\label{eqmutualen}
\varepsilon(\mathbf{z}) =- \frac{Gm}{2}\int \frac{\tilde \rho_{\rm int}(\mathbf{y})}{|\mathbf{z}-\mathbf{y}|}d^3\mathbf{y}
%
%\sum_j\int d^{3n-3} \mathbf{Y}'
%\frac{|\Psi_{\rm int}(t,\mathbf{Y}')|^2}
%{|\mathbf{z}  -\mathbf{y}_j' |}\,,
\end{align}
as {\it half} the gravitational potential energy of a mass $m$ at location $\mathcal{\mathbf{z}}$ (in CM frame), due to the entire lattice, and
\begin{equation}
\tilde \rho_{\rm int}(\mathbf{y}) = m\sum_{j=1}^n \int \delta(\mathbf{y}-\mathbf{y}_j') |\Psi_{\rm int}(\mathbf{Y}')|^2 d^{3n-3} \mathbf{Y}'
\end{equation}
is the CM-frame mass density. (Note: $\mathbf{y}_n \equiv -\sum_{j=1}^{n-1} \mathbf{y}_j$).  We will now have to show that $V_{\rm SN}$ approximately separates into a sum of terms that either only depend on $\mathbf{Y}$, or only on $\mathbf{x}$.  Taylor expansion of $V_{\rm SN}$  in $\mathbf{x}$ and $\mathbf{z}$ leads to (for one direction):
\begin{align}
\label{eqVSN}
V_{\rm SN} =& \sum_k \varepsilon(\mathbf{y}_k)+
 (x_{\rm CM} -\langle x_{\rm CM }\rangle )\sum_k\varepsilon'(\mathbf{y}_k)\nonumber\\
 +&
\frac{x^2_{\rm CM} -2x_{\rm CM}\langle x_{\rm CM}\rangle+\langle x_{\rm CM}^2\rangle}{2}\sum_k\varepsilon''(\mathbf{y}_k)  \,,
\end{align}
with higher orders fall as powers of $\Delta x_{\rm CM}/\Delta x_{\rm zp} \ll 1$.  Here in $V_{\rm SN}$, the first term  describes the leading SN correction to internal motion, and can be absorbed into $H_{\rm int}$. The second term describes the interaction between CM motion and each individual atom --- it can be shown to have negligible effects, because internal motions of different atoms are largely independent, and at much faster time scales. The third term is largely a correction to the CM motion; its main effect is captured if we replace it by its ensemble average over internal motion (again allowed by approximate independence between atoms, see Supplementary Material):
$
\sum_k\varepsilon''(\mathbf{y}_k) \rightarrow  \mathcal{C} \equiv  \Big\langle\sum_k\varepsilon''(\mathbf{y}_k) \Big\rangle$, with
%\,.
%\end{equation}
\begin{equation}
\label{eqC}
\mathcal{C}=- \frac{1}{2}\frac{\partial^2}{\partial z^2}\left[ \int \frac{G\tilde \rho_{\rm int}(\mathbf{y})\tilde\rho_{\rm int}(\mathbf{y}')}{|\mathbf{z} +\mathbf{y} -\mathbf{y}'|}d\mathbf{y} d\mathbf{y}' \right]_{\mathbf{z}=0}\,,
\end{equation}
which is half the double spatial derivative of the ``self-gravitational energy'' of the lattice as it is being translated.  As this is independent from the internal motion $\bf  Y$,  we therefore obtain the leading correction to $H_{\rm CM}$, which justifies Eq.~\eqref{eqreal} introduced at the beginning.

\noindent {\it Estimates for $\omega_{\rm SN}$.---}  Let us now estimate the magnitude of $\omega_{\rm SN}$ from Eq.~\eqref{eqC}.  Naively assuming a  homogeneous mass distribution with constant density $\rho_0$ leads to
\begin{equation}\label{eqSN0}
\mathcal{C}^{\rm hom} \approx  GM\rho_0, \;
\omega_{\rm SN}^{\rm hom} \approx \sqrt{G\rho_0},
\end{equation}
up to a geometric factor that depends on the shape of the object. This is a typical estimate for the gravity-decoherence time scale for a homogeneous object prepared in a nearly Gaussian quantum state with position uncertainty much less than its size~\cite{romero2}.   Using the mean density of Si crystal, this is roughly $4\times 10^{-4}\,\mathrm{s}^{-1}$.
However, mass in a lattice is {\it highly concentrated} near lattice sites; the realistic $\tilde\rho_{\rm int}$ at low temperatures contains a total mass of $m$ around each lattice point,  Gaussian distributed with uncertainty of $\Delta x_{\rm zp}$ in each direction.  This gives, through Eq.~\eqref{eqC},
\begin{equation}
\omega_{\rm SN}^{\rm crystal} =  \sqrt{Gm/(12\sqrt{\pi}\Delta x_{\rm zp}^3)}\,.
\end{equation}
For $\Delta x_{\rm zp} \approx 4.86 \times 10^{-12}$\,m, we obtain
$
\omega_{\rm SN}^{\rm Si} \approx 0.036\mathrm{s}^{-1}
$, nearly 100 times $\omega_{\rm SN}^{\rm hom}$. If we define
\begin{equation}
\label{eqlambda}
\Lambda =\big({\omega_{\rm SN}^{\rm crystal}}/{\omega_{\rm SN}^{\rm hom}}\big)^2= {m}/({12\sqrt{\pi}\rho_0\Delta x_{\rm zp}^3 })\,,
\end{equation}
then $\Lambda = 8.3\times 10^3$ for Si crystal.
%This has the similar origin as the M\"ossbauer Effect, which also arises from the fact that atoms in lattices are tightly bound to the close vicinity of lattice points.

\noindent {\it Evolutions of Gaussian States and Experimental Tests.---} As one can easily
prove, Gaussian states remain Gaussian under Eq.~(\ref{eqreal}); the self-contained evolution equations for first and second moments of $\hat x$ and $\hat p$, which completely determine the evolving Gaussian state, are given by:
\begin{align}\label{eqmulti}
{\langle \dot{\hat x} \rangle} &={\langle \hat p \rangle}/M ,\,\quad~~\, 
{\langle \dot{\hat p} \rangle} =-M\omega_{\rm CM}^2\langle \hat x \rangle\,,\\
\dot{V}_{xx}&= 2 V_{xp}/M\,,\quad
\dot{V}_{pp}=- 2M(\omega_{\rm CM}^2+\omega_{\rm SN}^2)V_{xp} \,,\\
\dot{V}_{xp}&= V_{pp}/M-M(\omega_{\rm CM}^2+\omega_{\rm SN}^2) V_{xx}\,.
\label{eqmultiend}
\end{align}
For covariance we have defined $V_{AB}\equiv \langle \hat A\hat B+\hat B\hat A \rangle/2 -\langle\hat A \rangle\langle \hat B\rangle$. Eq.~\eqref{eqmulti} indicates that  $\langle \hat x\rangle$  and $\langle\hat p\rangle$ evolve the same way as a harmonic oscillator with angular frequency $\omega_{\rm CM}$ --- any {\it semiclassical} measurement
of on  $\langle\hat x\rangle$ and $\langle\hat p\rangle$ will confirm classical physics. On the other hand,
evolution of second moments  (which represent {\it  quantum uncertainty}), is modified to that of a harmonic oscillator with a different frequency (see Fig.~\ref{fig:2freq}):
\begin{equation}
\omega_{\rm q} \equiv  \sqrt{\omega_{\rm CM}^2+\omega_{\rm SN}^2}\,.
\end{equation}
Equations~\eqref{eqmulti}--\eqref{eqmultiend} for Gaussian states can also be reproduced by a set of effective Heisenberg equations that contain expectation values:
 \begin{align}
 \dot{\hat x} = \hat p/M \,,\quad
 \dot{\hat p} =&-M\omega_{\rm CM}^2 \hat x - \mathcal{C}(\hat x-\langle \hat x\rangle)\,.
  \label{eqpdot}
 \end{align}
Classical gravity introduces a $\mathcal{C}$-dependent term to Eq.~\eqref{eqpdot}, in a way that only affects quantum uncertainty.

The most obvious test for the SN effect is to prepare a mechanical oscillator into a squeezed initial state, let it evolve for a duration $\tau$, and carry out state tomography.  We  need to detect an extra phase
$\Delta\theta =  \omega_{\rm CM}\tau ({ \omega_{\rm SN}^2}/{\omega_{\rm CM}^2})
$ in the rotation of the quantum uncertainty ellipse.  This seems rather difficult because $\omega_{\rm SN}/\omega_{\rm CM}$ is often a very small number, yet $\omega_{\rm CM} \tau$ is often not large, either.

However, we have not taken advantage of the fact that  $\Delta \theta$ is deterministic and repeatable.  One  way of doing so is to carry out a frequency-domain experiment.  Suppose we use light (at $\omega_0$) to continuously probe  a mechanical object's position, with quantum back-action noise (in the form of radiation-pressure noise) comparable in level to thermal noise, as has  been achieved by Purdy {\it et al.}~\cite{Purdy}. The effective Heisenberg equations (valid for Gaussian states) for such an optomechanical device is given by:
 \begin{align}\label{eqeom1}
 \dot{\hat x} =& \hat p/M\,, \\\label{eqeom2}
 \dot{\hat p} =&-M\omega_{\rm CM}^2 \hat x - 2\gamma_m \hat p -\mathcal{C}(\hat x-\langle \hat x\rangle)+\hat F_{\rm BA} + F_{\rm th}\\\label{eqeom3}
 \hat b_2 =& \hat a_2  + n_x+ (\alpha/\hbar) \hat x\,, \quad
 \hat b_1 =\hat a_1\,.
 \end{align}
Here $\gamma_m$ is the damping rate, $\alpha$ the optomechanical coupling constant and, $\hat F_{\rm BA}\equiv \alpha \,\hat a_1$ the quantum back action, and $F_{\rm th}$ the classical driving force (e.g., due to thermal noise). $\hat a_{1,2}$ represent quadratures of the in-going optical field, and $\hat b_{1,2}$ those of the out-going field.  (They correspond to amplitude and phase modulations of the carrier field at $\omega_0$.)  We have used $n_x$ to denote  sensing noise.  As we show in the Supplementary Material \cite{supplyc}, the out-going quadrature $\hat b_2$ contains two prominent frequency contents, peaked at $\omega_{\rm CM}$ (due to classical motion driven by thermal forces) and at $ \omega_{\rm q}$ (due to quantum motion driven by quantum fluctuation of light), respectively. Both have the same width ($\gamma_m$), and height (if thermal and back-action noises are comparable).  In order to distinguish them, we require
\begin{equation}
\label{eqQ}
S_{F_{\rm th}}\approx S_{F_{\rm BA}}\,,\quad Q \stackrel{>}{_\sim} ({\omega_{\rm CM}}/{\omega_{\rm SN}})^2\,
\end{equation}
This indicates a SN-induced shift of $\Delta \theta \approx 2\pi/Q$ per cycle can be picked up by the frequency domain experiment, even in presence of classical thermal noise $F_{\rm th}$.

For Si oscillators with $\omega_{\rm SN} \approx 0.036\,{\rm s}^{-1}$,  if $\omega_{\rm CM} \approx 2\pi \times 10\, {\rm Hz}$, Eq.~\eqref{eqQ} requires $Q \stackrel{>}{_\sim} 3\times 10^6$, which is challenging but possible~\cite{corbitt}. If a lower-frequency oscillator, e.g., a torsional pendulum with $\omega_{\rm CM} \approx 2\pi \times 0.1\,$Hz~\cite{toba}
can be probed with back-action noise above thermal noise, then we only require $Q\stackrel{>}{_\sim} 3\times 10^2$.

\noindent {\it SN equation for two macroscopic objects.---} Now suppose we have two objects confined within potential wells frequencies $\omega_{1,2}$, and moving along the same direction as the separation vector $\mathbf{L}$ connecting their equilibrium positions (from $1$ to $2$).  The standard approach for describing this interaction is to add a potential
\begin{equation}
\label{eqV12}
 V_g  = \mathcal{E}_{12}'  \big[ x_{{\rm CM}}^{(1)} - x_{{\rm CM}}^{(2)}\big]
+\left({\mathcal{C}_{12}}/2 \right) \big[ x_{{\rm CM}}^{(1)} -  x_{{\rm CM}}^{(2)}\big]
^2
\end{equation}
into the Schr\"odinger equation, with
\begin{equation}
\mathcal{E}_{12} \!\equiv \!- \int \! d^3 \mathbf{x} d^3\mathbf{y} \frac{G{ \tilde \rho}^{(1)}_{\rm tot}(\mathbf{x}) {\tilde \rho}_{\rm tot}^{(2)}(\mathbf{y})}{|{ { {\mathbf{L}+\mathbf{y}}-\mathbf{x}}}|},\;
\mathcal{C}_{12} \equiv \frac{\partial^2 \mathcal{E}_{12}}{\partial L^2},
\end{equation}
with ${\tilde \rho}_{\rm tot}^{(1)}$ and ${\tilde \rho}_{\rm tot}^{(2)}$ the mass densities of objects $1$ and $2$, respectively.
As has been argued by Feynman, this way of including gravity tacitly assumes that gravity is quantum. Although quantum operators have not been assigned for the gravitational field, they can be viewed as have been adiabatically eliminated due to their fast response: quantum information can transfer between these objects via gravity.   Suppose $\omega_1 =\omega_2 =\omega$, then $ V_g$  modifies the frequency of the two objects' differential mode  ---  allowing quantum state to slosh between them, at a frequency of
$\Delta =|\omega_+ -\omega_-| = \mathcal{C}_{12}/(2M\omega)$.
%In order for information to successfully transfer, one requires both objects to be in equilibrium with a heat bath with $k_B T \stackrel{<}{_\sim} {\hbar} \Omega_{\rm c}$, and $\Delta \stackrel{>}{_\sim} \omega_0/Q$, which requires
%\begin{equation}
%\label{Qinfo}
% Q  \stackrel{>}{_\sim} M \omega_0^2/ \mathcal{C}_{12}\,.
%\end{equation}

Suppose we instead use the SN equation for the two macroscopic objects.  In addition to modifying each object's own motion, we add a mutual term of
\begin{align}
\label{VSNmutual}
V_{\rm SN} =& \mathcal{E}'_{12}   \left[ x_{{\rm CM}}^{(1)} - x_{{\rm CM}}^{(2)}\right] \nonumber\\
+& \frac{\mathcal{C}_{12}}{2}
\left[
\big(x^{(1)}_{\rm CM}  - \langle x^{(2)}_{\rm CM}  \rangle\big)^2
+\big(x^{(2)}_{\rm CM}  - \langle x^{(1)}_{\rm CM}  \rangle\big)^2
\right].
\end{align}
This $V_{\rm SN}$   makes sure that only $\langle x_{\rm CM}\rangle$ gets transferred between the two objects the same way as in classical physics: {\it quantum uncertainty does not transfer from one object to the other.}  To see this more explicitly for Gaussian states, we can write down the full set of effective Heisenberg equations governing these two CMs:
\begin{align}
\dot{\hat x}_j = &\hat p_j/M_j\,, \nonumber\\
 \dot{\hat p}_j =& - M_j \omega_{\rm CM}^2 \hat x_j  -\sum_{k, j}\left[ \mathcal{E}_{kj}' +
 \mathcal{C}_{kj}\left(\hat x_j-\langle \hat x_k\rangle\right)\right]\,.
\end{align}
It is clear that expectation values follow classical physics, and quantum uncertainties are confined within each object --- and evolve with a shifted frequency.  Although we have shown theoretically that the inability of transferring quantum uncertainty and the shift between $\omega_{\rm CM}$ and $\omega_{\rm q}$  {\it share the same origin}, in practice, observing the frequency shift for a single object will be much easier, because  $\mathcal{C}_{12} \sim   G M^2/L^3 \stackrel{<}{_\sim} GM\rho_0 \ll \mathcal{C}_{11},\mathcal{C}_{22}$, due to the lack of the amplification factor $\Lambda$ in $\mathcal{C}_{12}$  [cf. Eq.~\eqref{eqlambda}].

\noindent {\it Discussions---} The lack of experimental tests on the quantum coherence of dynamical gravity makes us believe that semiclassical gravity is still worth testing~\cite{carlip}.  Our calculations have shown that signatures  of classical gravity in macroscopic quantum mechanics, although extremely weak, can be detectable with current technology. In particular, the classical self gravity of a single macroscopic object causes a much stronger signature than the classical mutual gravity between two separate objects: simply because the mass of a cold crystal is concentrated near lattice sites.   We also speculate that the rate of gravity decoherence should also be expedited by $\Lambda^{1/2} \sim 100$ --- if it is indeed determined by gravitational self energy~\cite{penrose1,penrose2}.  However, due to the lack of a widely-accepted microscopic model for gravity decoherence, this only makes it more hopeful for experimental attempts, but would not enforce a powerful bound if decoherence were not to be found.

Finally, since  classical gravity  requires the existence of a global wave function of the universe that does not collapse, (the unlikely case of) a positive experimental result will open up  new opportunities of investigating the  nature of quantum measurement.

\noindent {\it Acknowledgement---}We thank R.X.\ Adhikari, M.\ Aspelmeyer, C.M.\ Caves,  E.E.\ Flanagan, B.L.\ Hu, Y.\ Levin, H.\ Wiseman and our  colleagues in the LIGO MQM  group for fruitful discussions. We thank W.W.\ Johnson for pointing us to literature on the Debye-Waller factor. We acknowledge funding provided by the Institute for Quantum
Information and Matter, an NSF Physics Frontiers Center with support of
the Gordon and Betty Moore Foundation.  Y.C.\ thanks the Keck Institute for Space Studies for support. This work has also been supported by NSF
Grants Nos.\ PHY-0555406, PHY-0956189, PHY-1068881, and NSC Grant No.\ 100-2112-M-259-001-MY3 as well as the David and Barbara
Groce startup fund at Caltech.

\appendix

\section{Supplementary material for Macroscopic Quantum Mechanics in a Classical Spacetime}

\subsection{A. Incompatibility between the many-world interpretation of quantum mechanics and classical gravity}
\label{app:manyworld}

At this moment, the only well-known (and widely accepted) interpretation of quantum mechanics that explains the phenomenology of quantum measurement without resorting to quantum-state reduction is  the many-world interpretation of quantum mechanics [20]: the entire universe's wavefunction contains many branches, incorporating all possible measurement outcomes; each observer, however, can only perceive one of the branches, therefore experiencing the phenomenon of quantum-state reduction.

If we were to combine the many-world interpretation of quantum mechanics and classical gravity, the classical spacetime geometry will have to be determined by the expectation value of stress-energy tensor, which effectively averages over all possible measurement outcomes.   Following Page and Geilker [21], let us consider the following gedankenexperiment. Suppose quantum measurement of $\hat \sigma_z$ of a spin-1/2 particle at a state of $(|+\rangle + |-\rangle)/\sqrt{2}$ determines whether we put a mass on the left or right side of a scale, then the combination of the many-world interpretation and classical gravity will predict a leveled scale, because the expectation value of matter densities on both sides are equal.  This is in stark contrast with experimental facts.

Nevertheless, for many, including some of the authors, neither the concept of state reduction nor the many-world interpretation seems a satisfactory explanation of why we cannot predict the outcome of a quantum measurement.   We therefore remain open minded towards the possibility of further interpretations/clarifications/modifications of quantum mechanics and the quantum measurement process  that offer better explanations.  For us, the Schr\"odinger-Newton equation is not ruled out right away, and is therefore still worth testing.

\renewcommand{\theequation}{B.\arabic{equation}}
\setcounter{equation}{0}

\subsection{B.\ Separation between CM and internal degrees of freedom}

In Eq.~(11), we obtained the Schr\"odinger-Newton potential kept at quadratic order (in $\Delta x_{\rm CM}/\Delta x_{\rm zp} \ll 1$):
\begin{eqnarray}
V_{\rm SN} &=& \sum_k \varepsilon(\mathbf{y}_k)+
 (x_{\rm CM} -\langle x_{\rm CM }\rangle )\sum_k\varepsilon'(\mathbf{y}_k)\nonumber\\
 &+&
\frac{1}{2} (x^2_{\rm CM} -x_{\rm CM}\langle x_{\rm CM}\rangle+\langle x_{\rm CM}^2\rangle)\sum_k\varepsilon''(\mathbf{y}_k) \,.
\end{eqnarray}
As we shall argue, truncation at this order will give us the leading correction to CM motion, and it is also separable from corrections to the internal motion, therefore justifying the assumption of separability between the CM motion and
internal DOFs. Higher order terms, being suppressed by powers of $\Delta x/\Delta x_{\rm zp}$ are therefore negligible.

The first term
\begin{equation}
V_{\rm SN}  ^{(0)} =  \sum_k \varepsilon(\mathbf{y}_k)
\end{equation}
is readily absorbed into $H_{\rm int}$, which now becomes nonlinear. This is already the leading correction for the internal motion.  Let us calculate the modulus of its contribution,
\begin{eqnarray}
\|V_{\rm SN}  ^{(0)} \varphi\| &=& \left[\int \sum_{k,j} \varepsilon(\mathbf{y}_k) \varepsilon(\mathbf{y}_j) |\Psi_{\rm int} (\mathbf{Y})|^2 d^{3n-3}\mathbf{Y}\right]^{1/2}  \nonumber\\
&\approx & n \left(Gm^2/\Delta x_{\rm zp} \right)
\end{eqnarray}

The second term
\begin{equation}
V^{(1)}_{\rm SN} =
 (x_{\rm CM} -\langle x_{\rm CM}\rangle)\sum_k\varepsilon'(\mathbf{y}_k)
\end{equation}
 describes the interaction between the CM and the internal motion of each of the atoms. In order to estimate its effect, let us first calculate the modulus of the change in $\varphi$ it induces:
\begin{align}
\label{eqVSN1}
\| V_{\rm SN}^{(1)} \varphi \|
=& \Delta x_{\rm CM} \left[\int\sum_{k,j} \varepsilon'(\mathbf{y}_k) \varepsilon'(\mathbf{y}_j) |\Psi_{\rm int} (\mathbf{Y})|^2 d^{3n-3}\mathbf{Y}\right]^{1/2} \nonumber\\
\approx & \sqrt{n} \left( {G m^2}/\Delta x_{\rm zp}\right) \left(\Delta x_{\rm CM}/\Delta x_{\rm zp}\right)\,.
\end{align}
Here we only have $\sqrt{n}$ (instead of $n$) because the integral will not vanish only when distribution of $\mathbf{y}_k$ and $\mathbf{y}_j$ are correlated, which only happens for nearby atoms.
 The other factor $\Delta x_{\rm CM}/\Delta x_{\rm zp}$ is due to the fact that this is the next order in the Taylor expansion.  From the point of view of internal motion, $V_{\rm SN}^{(1)}$ clearly gives a higher-order correction than $V_{\rm SN}^{(0)}$, hence negligible.  We will show that, even for CM motion,  the contribution of $V_{\rm SN}^{(1)}$ is also less than contribution from the next Taylor-expansion term  $V_{\rm SN}^{(2)}$.

Now turning to $V_{\rm SN}^{(2)}$, let us split it into two terms
\begin{equation}
V_{\rm SN} ^{(2)} = \bar V_{\rm SN}^{(2)}  +
\delta V_{\rm SN} ^{(2)},
\end{equation}
with
\begin{equation}
\bar V_{\rm SN}^{(2)} =
\frac{1}{2} (x^2_{\rm CM} -2x_{\rm CM}\langle x_{\rm CM}\rangle+\langle x_{\rm CM}^2\rangle) \Big\langle \sum_k\varepsilon''(\mathbf{y}_k)\Big\rangle \,,
\end{equation}
defined as the ensemble average, where
\begin{equation}
\Big\langle \sum_k\varepsilon''(\mathbf{y}_k)\Big\rangle  \equiv \int \sum_k\varepsilon''(\mathbf{y}_k) |\Psi_{\rm int}(\mathbf{Y})|^2 d^{3n-3}\mathbf{Y} \equiv\mathcal{C}\,,
\end{equation}
and
\begin{equation}
\mathcal{C} =- \frac{1}{2}\frac{\partial^2}{\partial z^2}\left[ \int \frac{G\tilde \rho_{\rm int}(\mathbf{y})\tilde\rho_{\rm int}(\mathbf{y}')}{|\mathbf{z} +\mathbf{y} -\mathbf{y}'|}d\mathbf{y} d\mathbf{y}' \right]_{\mathbf{z}=0}\,.
\end{equation}

Note that $\bar V_{\rm SN}^{(2)}$ does not depend explicitly on $\mathbf{Y}$, and hence is a correction to the Hamiltonian for the CM motion.  It is straightforward to estimate that
\begin{equation}
\|  \bar V_{\rm SN}^{(2)} \varphi \| = n Gm^2/\Delta x_{\rm zp} \left(\Delta x_{\rm CM} /\Delta x_{\rm zp}\right)^2\,.
\end{equation}
This means, at any given time,
\begin{equation}
\|   V_{\rm SN}^{(1)} \varphi \| /\|  \bar V_{\rm SN}^{(2)} \varphi \| \approx \frac{1}{\sqrt{n}}\frac{\Delta x_{\rm zp}}{\Delta x_{\rm CM}} \approx \sqrt{\frac{\omega_{\rm CM}}{\omega_D}} \ll 1\,.
\end{equation}
In addition, as we evolve in time, the effect of $V_{\rm SN}^{(1)}$ oscillates around zero over a very fast time scale, while the effect of $\bar V_{\rm SN}^{(2)}$ does not oscillate around zero --- this further suppresses the relative contribution of $V_{\rm SN}^{(1)}$.  For this reason, we ignore $V_{\rm SN}^{(1)}$ completely.

As for $\delta V_{\rm SN}^{(2)}$, its effect is suppressed from $\bar V_{\rm SN}^{(2)}$ by $\sqrt{n}$, because, much similar to Eq.~\eqref{eqVSN1}, effects of different atoms do not accumulate unless they are very close to each other.

\renewcommand{\theequation}{C.\arabic{equation}}
\setcounter{equation}{0}

\subsection{C. Effective Heisenberg Equations of Motion and Coupling with Optical Field}

 The fact that Gaussian states leads to Gaussian states encourages us to look for {\it effective} Heisenberg equations of motion, which will at least be valid for Gaussian states.  It is easy to find that
 \begin{align}
 \dot{\hat x} =& \hat p/M \\
 \dot{\hat p} =&-M\omega_{\rm CM}^2 \hat x -{\mathcal C}(\hat x-\langle \hat x\rangle)
 \end{align}
 will give the same set of first- and second-moment equations of motion as the SN equation. Note that in the Heisenberg picture, the initial state of the oscillator remains constant.

Let us now consider a more realistic scenario, in which the oscillator is damped with decay rate $\gamma_m$, driven with classical thermal noise and other classical driving; we also consider using light to sense the position of the mirror, in which we also suffer from sensing noise.  The entire process can be described by the following set of equations [cf. Eqs.~(21), (22) and (23)]:
 \begin{align}
\dot{\hat x} =& \hat p/M\,, \\
 \dot{\hat p} =&-M\omega_{\rm CM}^2 \hat x- 2\gamma_m \hat p -{\mathcal C}(\hat x-\langle \hat x\rangle)+ {\hat F}_{\rm BA}+ F_{\rm th} \,,\\
 \hat b_2 =& \hat a_2  + n_x+ (\alpha/\hbar) \hat x\,, \quad
 \hat b_1 =\hat a_1\,.
 \end{align}

This set of equations can be solved first for $\langle \hat x\rangle$ by taking the expectation value of the first two equations, and then insert this back to obtain the entire solution.  If we define
\begin{align}
\chi_0& = -\frac{1}{M(\omega^2 +2i\gamma_m\omega -\omega_{\rm  CM}^2)} \\
\chi_g &= -\frac{1}{M(\omega^2 +2i\gamma_m\omega -\omega_{\rm q}^2)}\,,
\end{align}
then we can write, in the frequency domain,
\begin{equation}
\label{b2out}
\hat b_2 = \hat a_2 +(\alpha/\hbar)\chi_g \hat F_{\rm BA}+(\alpha/\hbar)\chi_0 F_{\rm th} +n_x\,.
\end{equation}
Because both $\chi_g$ and $\chi_0$ in the time domain are Green functions of stable systems, Eq.~\eqref{b2out} represent the steady-state solution for the out-going field, which is only determined by the in-going optical field and the classical driving field --- initial states of the mechanical oscillator does not matter (similar to the case of Ref.~[49, 50]).

Equation~\eqref{b2out} carries the separation between classical and quantum rotation frequencies in the previous section (Fig.~1) into the frequency spectrum of our measuring device: quantum back-action (radiation-pressure) noise $\hat F_{\rm BA}$ spectrum in the output port of the continuous measuring device is the same as an oscillator with frequency $\omega_{\rm q}$, and therefore peaks around $\omega_{\rm q}$ --- while classical noise $F_{\rm th}$ follows that of an oscillator with frequency $\omega_{\rm CM}$, and peaks at $\omega_{\rm CM}$.   In order to look for such a signature, we will need classical force noise to be comparable in level to quantum noise, and have the two peaks to be resolvable,
\begin{equation}
\omega_{\rm q} -\omega_{\rm CM} \stackrel{>}{_\sim} \gamma_m
\end{equation}
which means
\begin{equation}
Q\stackrel{>}{_\sim} (\omega_{\rm CM}/\omega_{\rm SN})^2
\end{equation}
where $Q$ is the quality factor of the mechanical oscillator.

\renewcommand{\theequation}{D.\arabic{equation}}
\setcounter{equation}{0}

\subsection{D.  SN Equation for Two Macroscopic Objects}

Following the analysis in Appendix.\ A, here we deal with two macroscopic objects, and define
\begin{align}
\mathbf{x}_k^{(I)} &=  \mathbf{X}^{(I)}+ \mathbf{y}_k^{(I)} +\mathbf{x}_{\rm CM}^{(I)}\,,\quad k=1,\ldots,n_I\,,\quad I=1,2.
\end{align}
Here $\mathbf{X}^{(I)}$ is the zero point we use for describing object $I$, and $n_I$ is the number of atoms object $I$ contains.  Following the same argument for deriving the single-object SN equation, we can still write the joint wavefunction as a product between the joint CM wavefunction and the internal wavefunctions,
\begin{equation}
\label{multivarphi}
\varphi =\Psi_{\rm CM}\left[\mathbf{x}^{(1)},\mathbf{x}^{(2)}\right] \Psi_{\rm int}^{(1)} \left[\mathbf{Y}^{(1)}\right]
\Psi_{\rm int}^{(2)} \left[\mathbf{Y}^{(2)}\right]\,.
\end{equation}
and show that this form will be preserved during evolution, even adding the SN term, which is now
\begin{widetext}
\begin{equation}
V_{\rm SN} = \frac{Gm^2}{2} \sum_{I,J=1}^2\sum_{i=1}^{n_I}\sum_{j=1}^{n_J} \frac{\left|\Psi_{\rm CM}(\tilde{\mathbf{z}}^{(1)},\tilde{\mathbf{z}}^{(2)}) \Psi_{\rm int}^{(1)} (\tilde{\mathbf{Y}}^{(1)}) \Psi_{\rm int}^{(2)} (\tilde{\mathbf{Y}}^{(2)})\right|^2 }
{\left|\mathbf{L}^{(JI)} + \mathbf{x}^{(I)} + \mathbf{y}^{(I)}_i - \tilde{\mathbf{z}}^{(J)} + \tilde{\mathbf{y}}^{(J)}_j \right|}
d\tilde{\mathbf{z}}^{(1)} d\tilde{\mathbf{z}}^{(2)} d\tilde{\mathbf{Y}}^{(1)} d\tilde{\mathbf{Y}}^{(2)}
\end{equation}
\end{widetext}
Here we have denoted
\begin{equation}
L^{(IJ)} \equiv X^{(J)} - X^{(I)}\,.
\end{equation}
Terms with $I=J$ have already been dealt with, and gives rise to the SN correction within object $I$.  We will only have to deal with cross terms.  In doing so, we shall assume each object's CM moves very little from its zero point, and carry out Taylor expansion.  Note that because these objects are already macroscopically separated, with $L^{(IJ)}$ comparable to or greater than the size of each object, the expansion here will be valid for the cross term as long as the CM motion of each object is much less than its size.

The zeroth order expansion in CM motion, $V_{\rm SN}^{(0)}$  gives rise to SN coupling between the objects' internal motions.  Fortunately, that does not entangle their internal motions, and preserves the form of Eq.~\eqref{multivarphi}.

The first order in CM motion gives (after conversion of summation over atoms into ensemble average, the same as we did in Appendix B, and removing a constant):
\begin{align}
\bar V_{\rm SN}^{(1)} = - x^{(1)}_{\rm CM} \mathcal{E}'_{21}
-   x^{(2)}_{\rm CM}  \mathcal{E}'_{12}
=  \left[x^{(1)}_{\rm CM} - x^{(2)}_{\rm CM}\right] \mathcal{E}'_{12}
\end{align}
where $\mathcal{E}_{12}$ is the interaction energy between the objects, as a function of their separation
\begin{equation}
\mathcal{E}_{12} (\mathbf{x}) \equiv
-\int d^3\mathbf{y}d^3\mathbf{z}
 \frac{{G}\tilde \varrho_{\rm int}^{(1)}(\mathbf{y})   \,\tilde \varrho_{\rm int}^{(2)}(\mathbf{\mathbf{z})}}{|\mathbf{z}+ \mathbf{x}-\mathbf{y}  + \mathbf{L}^{(12)}|}
\end{equation}
This describes the tendency of these objects to fall into each other. Similarly, we obtain the second order, which gives (apart from a constant)
\begin{align}
\bar V_{\rm SN}^{(2)} =\frac{\mathcal{E}_{12}''}{2}
\left[
\left(x^{(1)}_{\rm CM}  - \langle x^{(2)}_{\rm CM}  \rangle\right)^2
+\left(x^{(2)}_{\rm CM}  - \langle x^{(1)}_{\rm CM}  \rangle\right)^2
\right]\,,
\end{align}
which justifies Eq.~(27).


\begin{thebibliography}{9}
\bibitem{Bonifacio} P.\ M.\ Bonifacio {\it et al.},  Class.\ Quantum Grav.\ {\bf 26}, 145013  (2009).
%semiclassical quantizations
%
%
%
\bibitem{diosi2} L.\ Di\'osi L, Phys.\ Lett.\ A {\bf 120}, 377 (1987).
% Gravity decoherence
%
%
\bibitem{diosi3} L.\ Di\'osi, \pra {\bf 40}, 1165 (1989).
\bibitem{diosi4} L.\ Di\'osi and J.\ J.\ Halliwell, \prl {\bf 81},  2846 (1998).
\bibitem{penrose1} R.\ Penrose, Gen. Rel. Grav. {\bf 28}, 581 (1996)
\bibitem{penrose2} R.\ Penrose, {\it The road to reality: A complete guide to the laws of the universe,} Alfred A.\ Knopf inc., New York (2005).
%
%
\bibitem{Marshall} W. Marshall, C. Simon, R. Penrose, and D. Bouwmeester, Phys. Rev. Lett. {\bf 91}, 130401 (2003).
\bibitem{Hong} T.\ Hong {\it et al.}, arXiv:1110.3349 [quant-ph] (2011).
\bibitem{jasper} J. van Wezel and T.\ H.\ Oosterkamp, Proc.\ R.\ Soc.\ A {\bf 468}, 35 (2012).
%
%
%
%
\bibitem{kaltenbaek} R.\ Kaltenbaek {\it et al.}, Exp.\ Astron.\  {\bf 34}, 123 (2012).
\bibitem{romero1} O.\ Romero-Isart {\it et al.}, \prl {\bf 107}, 020405 (2011).
\bibitem{romero2} O.\ Romero-Isart, \pra {\bf 84}, 052121 (2011).

\bibitem{maggiore} M.\ Maggiore, Phys.\ Lett.\ B, {\bf 319}, 83 (1993).
\bibitem{das} S.\ Das and E.\ C.\ Vagenas, \prl {\bf 101}, 221301 (2008).
\bibitem{pikovski} I.\ Pikovski {\it et al.}, Nature Phys.\ {\bf 8}, 393 (2012).

\bibitem{moller}C.\ Moller, {\it Les Theories Relativistes de la Gravitation} Colloques Internationaux CNRX 91 ed A Lichnerowicz and M-A Tonnelat (Paris: CNRS) (1962).
\bibitem{rosenfeld} L.\ Rosenfeld, Nucl. Phys. {\bf 40}, 353 (1963).
\bibitem{carlip} S.\ Carlip, Class.\ Quantum Grav.\ {\bf 25}, 154010 (2008).

%
\bibitem{Wald} R.\ M.\ Wald, {\it General Relativity}, University of Chicago Press (1984).
%
\bibitem{Everett} H.\ Everett III, \rmp {29}, 454 (1957).
%
\bibitem{page} D.\ N.\ Page and C.\ D.\ Geilker, \prl {\bf 47}, 979 (1981).
%

%
\bibitem{diosi1} L.\ Di\'osi, Phys.\ Lett.\ A {\bf 105}, 199 (1984).
%
\bibitem{Moroz} I.\ M. Moroz, R.\ Penrose, and P.\ Tod, Class.\ Quantum Grav.\ {\bf 15}, 2733 (1999).
%
\bibitem{Harrison} R.\ Harrison, I.\ Moroz, and K.\ P.\ Tod,  Nonlinearity {\bf 16}, 101 (2002).
%
\bibitem{salzman}  P.\ J.\ Salzman and S.\ Carlip, arXiv:gr-qc/0606120 (2006).
%
%
\bibitem{adler} S.\ L.\ Adler, J.\ Phys.\ A {\bf 40}, 755 (2007).
\bibitem{Guzman} F.\ S.\ Guzm\'an and L.\ A.\ Urena-L\'opez. \prd  {\bf 69}, 124033 (2004).
%
\bibitem{meter} J.\ R.\ Meter, Class. Quant. Grav. {\bf 28}, 215013, (2011). In this article, the authors assumed that the CM degree of freedom satisfies the single-particle SN equation. This is significanly different from what we have obtained. In particular, in the limiting case when the CM wave-function spread approaches zero, their SN term diverges due to divergence in the self-gravitational energy of a single particle, and the state changes dramtically. By contrast, our SN term also approaches zero in this case, as it arises from the change in many-particle gravitional energy induced by the CM motion, which decreases as the uncertainty in the CM motion decreases.
%
%
%
\bibitem{optomechanics_review} See, e.g., recent review article by
M. Aspelmeyer, P. Meystre, and K. Schwab, Physics Today {\bf 65}, 29 (2012).
%
\bibitem{oconnel} A.\ D. O'Connel {\it et al.}, Nature {\bf 464}, 697 (2010);
\bibitem{teufel} J.\ D.\ Teufel {\it et al.}, Nature {\bf 475}, 359 (2011);
\bibitem{qfproduct}J. Chan {\it et al.}, Nature {\bf 478}, 89 (2011).
\bibitem{Purdy} T.\ P.\ Purdy,  R.W.\ Peterson, and C.\ A.\ Regal., arXiv:1209.6334 (2012).
%
\bibitem{BK} V.\ B.\ Braginsky and F.\ Ya.\ Khalili, {\it Quantum Measurement}, Cambridge University Press (1992).
%
\bibitem{coating_thermal_noise}
Y.\ Levin, Phys.\ Rev.\ D {\bf 57}, 659 (1998);  T.\ Hong, {\it et al.}, arXiv:1207:6145 (2012).
\bibitem{helge} H.\ M\"uller-Ebhardt {\it et al.}, {\pra} {\bf 80}, 043802 (2009).
H.\ M\"uller-Ebhardt, H.\ Rehbein, R.\ Schnabel, K.\ Danzmann, and Y.\ Chen , Phys.\ Rev.\ Lett.\ {\bf 100}, 013601 (2008);  F.\ Khalili {\it et al.}, Phys.\ Rev.\ Lett.\ {\bf 105}, 070403 (2010);
\bibitem{haixing} H.\ Miao {\it et al.}, {\pra} {\bf 81}, 012114 (2010).
%
%
\bibitem{thomas} B.\ Abbott et al, New. J. Phys. {\bf 11}, 073032 (2009);
A.\ R.\ Neben {\it et al.}, New.\ J.\ Phys.\ {\bf 14}, 115008 (2012).
%
\bibitem{LIGO3} H.\ Miao {\it et al.}, {\it Comparison of Quantum Limits in Interferometer Topologies for 3rd Generation LIGO}, LIGO Document T1200008 (2012); S.\ L.\ Danilishin and F.\ Ya.\ Khalili, Living Rev.\ Relativity {\bf 15}, 5 (2012).
%
\bibitem{supplyc} See section C of the Supplementary Material 
%
\bibitem{corbitt} T.\ Corbitt {\it et al.}, \prl {\bf 98}, 150802 (2007); Phys.\ Rev.\ Lett.\ {\bf 99}, 160801 (2007); Phys.\ Rev.\ A {\bf 73},  023801 (2006).
%
\bibitem{AEI}  T.\ Westphal {\it et al.}, App.\ Phys.\ B, 1-7  (2012); S.\ Gossler, Class. Quantum Grav. {\bf 27}, 084023  (2010).
%
\bibitem{CLIO} T.\ Uchiyama {\it et al.}, \prl {\bf 108}, 141101  (2012).
%
%
\bibitem{debye} C. Kittel, {\it Introduction to Solid State Physics}, 8th ed. {\it Wiley} (2004).
\bibitem{BF} M.\ Blackman and R.\ H.\ Fowler, Math.\ Proc.\ Cambridge Phil.\ Soc.\ {\bf 33}, 380 (1937).
%
\bibitem{DWF} R.\ M.\ Housley and F.\ Hess, Phys.\ Rev.\  {\bf 146}, 517  (1966).
%
\bibitem{Flensburg} C.\ Flensburg and R.\ F. Stewart. \prb {\bf 60}, 284 (1999).
%
%
%
%\bibitem{NV} D.\ Marcos {\it et al.}, \prl {\bf 105}, 210501 (2010); L.\ Childress {\it et al.}, Science {\bf 314}, 281 (2006); T.\ Togan {\it et al.}, Nature (London ) {\bf 466}, 730 (2010)


\bibitem{toba} K.\ Ishidoshiro {\it et al.}, \prl {\bf 106}, 161101 (2011).
\bibitem{BGKMTV} V.\ B.\ Braginsky {\it et al.}, \prd {\bf 67}, 082001 (2003).
\bibitem{Khalili2012} F.\ Ya.\ Khalili {\it et al.}, Phys.\ Rev.\ A {\bf 86}, 033840 (2012).
\end{thebibliography}
\end{document}